\documentclass[12pt]{article}
\usepackage{amsfonts}
\usepackage{amssymb}
\usepackage{amsmath}
\usepackage{amsthm}
\usepackage{graphicx}

\begin{document}

\title{\bf On singularity-resolution in mimetic gravity}

\author{\Large Suddhasattwa Brahma${}^a$, Alexey Golovnev${}^{a,b}$ and Dong-han Yeom${}^{a,c}$\\
${}^{a}${\it Asia Pacific Center for Theoretical Physics,}\\ 
{\it 67 Cheongam-ro, Nam-gu, Pohang 37673, Republic of Korea}\\
${}^{b}${\it Faculty of Physics, St. Petersburg State University,}\\ 
{\it Ulyanovskaya ul., d. 1, Saint Petersburg 198504, Russia}\\
${}^{c}${\it Department of Physics, POSTECH,}\\ 
{\it 67 Cheongam-ro, Nam-gu, Pohang 37673, Republic of Korea}\\
{\small suddhasattwa.brahma@gmail.com \quad agolovnev@yandex.ru \quad  innocent.yeom@gmail.com}}
\date{}

\maketitle

\begin{abstract}
Recently, it was shown that modified mimetic gravity, with a $f(\square\phi)$ term, results in a singularity-free model of gravity, for both cosmological and black hole spacetimes \cite{CMcosmo,CMBH}. In this work, we analyze this model further and show that, since the function $f$ was tuned to vanish rapidly for small values of the argument, the non-singular bounce relies heavily on a subtle branch changing mechanism for the multi-valued function $f$. Furthermore, this mechanism has interesting implications for the proposed link between this model and loop quantum cosmology.
\end{abstract}

\section{Introduction}
Notoriously, our best description of gravitational interactions, the theory of General Relativity (GR), generically leads to singularities as long as reasonable energy conditions for matter fields are satisfied \cite{Singularity}. The singular loci are usually assumed to indicate limitations of the classical theory. And indeed, in high curvature regions it should  anyway be superseded by a quantum theory. However, the problem is exacerbated by the fact that a full theory of quantum gravity is nowhere close to being at hand.

Much effort has been directed at finding a model of gravity which would ameliorate the issue of singularities. One of the very interesting recent attempts appeared in Refs. \cite{CMcosmo, CMBH}. It builds upon the model of mimetic gravity \cite{CMmimetic} which amounts to a modification of the conformal sector in a way that is equivalent \cite{mymimetic} to introducing a scalar field $\phi$ subject to the constraint $(\partial\phi)^2=-1$. The crucial new ingredient is a non-linear function $f(\square\phi)$ added to the Lagrangian density which is supposed \cite{CMcosmo, CMBH} to begin only from the fourth order in Taylor expansion around $\square\phi=0$ for avoiding unwanted corrections to the weak gravity regime, the latter being  tightly constrained experimentally.

It was argued that both cosmological \cite{CMcosmo} and black hole \cite{CMBH} singularities can be resolved. The physical idea behind this proposal is remarkably simple. In a synchronous coordinate system $ds^2=-dt^2+\gamma_{ij}dx^i dx^j$, with the scalar field $\phi=\pm t+const$, we have $\square\phi=\mp\frac{\dot\gamma}{2\gamma}$, and therefore approaching the singularity features an unbounded increase of the argument for the function $f$. Choosing some limiting value $\chi_m$ for $\chi\equiv\square\phi$, preferably well below the Planck scale in order to avoid quantum gravitational features\footnote{This requirement shall be relaxed later in order to make contact with loop quantum cosmology (LQC).}, we can adjust \cite{CMcosmo, CMBH} for a Born-Infeld type behaviour of $f$ when $\chi\to\chi_m$. It was demonstrated that it leads to a bounce-like resolution of singularities.

Here we further the analyses and wish to report that this mechanism heavily relies on use of a multi-valued function $f(\square\phi)$ with a subtle mechanism for changing the branch (see also Ref. \cite{HP}). The intuitive reason lies in the fact that $\square\phi$ is not a good measure of spacetime curvature. In particular, for a Friedmann universe we have $\square\phi=-3H$ where $H$ is the Hubble parameter while the scalar curvature also depends on $\dot H$. By definition, a bounce corresponds to $H=0$ which is precisely the zero value of the argument for the function $f(\chi)$ which has been tuned to vanish faster than $\chi^3$ around zero. Therefore, in the absence of a branch change, at the posited bounce the model is equivalent to GR, and no bounce is possible in GR.

Below we  explain some details of the Chamseddine-Mukhanov (CM) model (Sec.~\ref{sec:the}), describe the mechanism of changing the branch (Sec.~\ref{sec:nob}, Sec.~\ref{sec:dyn}, and Sec.~\ref{sec:bey}), and discuss implications for previously asserted relations to LQC (Sec.~\ref{sec:ons}).

\section{\label{sec:the}The modified mimetic gravity}
The action of the CM model is of the form \cite{CMcosmo, CMBH} 
\begin{equation}
\label{action}
S=\int d^4 x\sqrt{-g}\cdot\left(\frac12 R-\lambda\left(1+(\partial\phi)^2\right)-f(\square\phi)\right)
\end{equation}
where the function $f$ was taken to be
\begin{equation}
\label{f}
f(\chi)=-\chi_m^2\left(1+\frac{\chi^2}{3\chi_m^2}-\frac{\sqrt{2}\chi}{\sqrt{3}\chi_m}\arcsin\left(\frac{\sqrt{2}\chi}{\sqrt{3}\chi_m}\right)-\sqrt{1-\frac{2\chi^2}{3\chi_m^2}}\right)
\end{equation}
with 
\begin{equation}
\label{condition}
f(0)=f^{\prime}(0)=f^{\prime\prime}(0)=0
\end{equation} 
and $f^{\prime\prime\prime}(0)=0$, too. Note that the sign of $f$ is changed compared to Ref. \cite{CMcosmo} due to the opposite sign convention for the metric signature without changing the relative sign between $R$ and $f$ terms in the action. The zero values of the function and some of its derivatives, at $\chi=0$, correspond to the principal branch and results in the theory matching with GR in the weak-curvature limit.

It is not difficult to perform the variations in the action (\ref{action}) which give the Einstein equation ($\delta g^{\mu\nu}$)
\begin{equation}
\label{Einstein}
G_{\mu\nu}=2\lambda\phi_{,\mu}\phi_{,\nu}-\left(\phi_{,\mu}f^{\prime}_{,\nu}+\phi_{,\nu}f^{\prime}_{,\mu}\right)+g_{\mu\nu}\left(\phi^{,\alpha}f^{\prime}_{,\alpha}-f+f^{\prime}\square\phi\right)
\end{equation}
with the usual Einstein tensor $G_{\mu\nu}$, the constraint ($\delta\lambda$)
\begin{equation}
\label{constraint}
(\partial\phi)^2\equiv g^{\mu\nu}\phi_{,\mu}\phi_{,\nu}=-1,
\end{equation}
and the wave equation ($\delta\phi$)
\begin{equation}
\label{wave}
\square f^{\prime}=2\bigtriangledown_{\mu}\left(\lambda\partial^{\mu}\phi\right).
\end{equation}

\section{\label{sec:nob}No bounce in trivial branch}
The simplest situation in which avoidance of singularity was claimed \cite{CMcosmo} is the spatially flat Friedmann universe
\begin{equation}
\label{Friedmann}
ds^2=-dt^2+a^2(t)dx^2
\end{equation}
with $\phi=\phi(t)$, and therefore $\phi=\pm t+ const$. We now proceed to show that no bounce is possible for any single-valued function $f$ satisfying the condition (\ref{condition}).

We easily find that $$\square\phi=\mp 3H$$ where $H\equiv \frac{\dot a}{a}$ is the Hubble constant and $f^{\prime}_{,\mu}=\mp 3f^{\prime\prime}{\dot H}\delta^0_{\mu}$. Substituting it together with the metric ansatz (\ref{Friedmann}) into the Einstein equation (\ref{Einstein}), we get
\begin{equation}
\label{Fried1}
3H^2=2\lambda+3f^{\prime\prime}{\dot H}+f+3Hf^{\prime}
\end{equation}
for the temporal component, and
\begin{equation}
\label{Fried2}
2{\dot H}+3H^2=-3f^{\prime\prime}{\dot H}+f+3Hf^{\prime},
\end{equation}
for the spatial ones.
The wave equation (\ref{wave}) reduces to
\begin{equation}
\label{redundant}
3f^{\prime\prime}{\ddot H}+9H{\dot H}f^{\prime\prime}-9{\dot H}^2f^{\prime\prime\prime}+2{\dot\lambda}+6H\lambda=0.
\end{equation}

It should not come as a surprise that equation (\ref{redundant}) has $\ddot H$, involving the third time derivative of the metric. We have a higher derivative action for $\phi$ which must produce fourth-order equations. However, the field $\phi$ is constrained (\ref{constraint}), implying that its first derivatives are expressed in terms of metric components without derivatives. Substituting this solution back, we get third-order equation for the metric. It might seem dangerous at first sight, but it is not a problem we report. Note that in Ref. \cite{Vikman} an equivalent action without higher derivatives is given for a similar model.

Actually, we can check that the Bianchi identity holds which renders equation (\ref{redundant}) redundant. Indeed, subtracting the Friedmann equations (\ref{Fried1}) and (\ref{Fried2}) from each other we see $2{\dot H}=-2\lambda-6f^{\prime\prime}{\dot H}$. After that we differentiate the first equation (\ref{Fried1}) with respect to time and substitute the above expression for $2\dot H$ into the $6H\dot H$ term. The final result is identical to the equation of motion (\ref{redundant}). 

Now let us come back to our statement that the bounce is not possible for the CM model in the trivial branch. Indeed, since at the bounce point $H=0$ the function $f$ is supposed to satisfy the condition (\ref{condition}), or $f(0)=f^{\prime}(0)=f^{\prime\prime}(0)=0$, we see that $H=0$ implies $\dot H=0$ by virtue of equation (\ref{Fried2}). It readily shows that the solution $a=\left(1+\frac34 \epsilon_m t^2\right)^{1/3}$ from the Ref. \cite{CMcosmo} cannot be true under these assumptions. In the next section, we will see how the model actually works, crucially depending on a subtle change of branch of the function $f$. 

Let us end this section by noting that adding matter would not help much since it gives a contribution of $+\rho$ to the r.h.s. of eq. (\ref{Fried1}) and $-p$ to the r.h.s. of eq. (\ref{Fried2}) with $\rho$ and $p$ being its energy density and pressure respectively. Then at the bounce point, an unacceptably pathological violation of energy conditions is required, precisely the same as in GR.\footnote{Let us mention for completeness that of course a trivial bounce is possible. One has to take some normal matter and mimetic fluid with negative density. Then formally it can work. An effective antigravitating fluid anyway lacks motivation and of course, its presence would result in a bounce even without the function $f$ in the action.}

\section{\label{sec:dyn}Dynamical change of branch}\label{BranchChange}
The argument in the Ref. \cite{CMcosmo} goes in the following way. The wave equation (\ref{wave}) can be written down as 
$$\frac{d}{dt}\left(2\sqrt{\gamma}\lambda\right)=\frac{d}{dt}\left(\sqrt{\gamma}f^{\prime\prime}\dot\chi\right)$$
and has the obvious solution 
\begin{equation}
\label{lambdasol}
\lambda=\frac{C}{2\sqrt{\gamma}}+\frac12 f^{\prime\prime}\dot\chi
\end{equation}
with an integration constant $C$ which corresponds to the density of mimetic fluid. Substituting it to the first Friedmann equation (\ref{Fried1}) and performing some algebra, we bring the modified Friedmann equation into a very simple form \cite{CMcosmo}:
\begin{equation}
\label{modFried}
3H^2=\epsilon\left(1-\frac{\epsilon}{\epsilon_m}\right)
\end{equation}
with the effective energy density given by $$\epsilon=\frac{C}{\sqrt{\gamma}}=\frac{C}{a^3}.$$ 
Clearly, such an equation as (\ref{modFried}), rather well-explored in LQC, does have a bouncing solution.

But can one arrive at a bouncing solution, as derived in \cite{CMcosmo}, despite the condition (\ref{condition}) which, according to the Friedmann equation (\ref{Fried1}), requires to set $C=0$ leaving Minkowski space as the only option? The answer is in the affirmative with the details of the claim as follows. For simplicity of presentation we choose $\chi_m=\sqrt{\frac23}$ in the function (\ref{f}):
\begin{equation}
\label{rf}
f(\chi)=-\frac23 \left(1+\frac12 \chi^2-\chi \arcsin\chi- \sqrt{1-\chi^2}\right).
\end{equation}

Given eq. (\ref{rf}), the Raychaudhuri equation (\ref{Fried2}) takes the form
\begin{equation}
\label{rRay}
-\frac23 \dot\chi+\frac13 \chi^2=\dot\chi\left(1-\frac{1}{\sqrt{1-\chi^2}}\right)+1-\frac12 \chi^2-\sqrt{1-\chi^2}.
\end{equation}
The cosmological solution from the Ref. \cite{CMcosmo} for this value of $\chi_m$ is very simple:
\begin{equation}
\label{bounce}
a=\left(1+ t^2\right)^{1/3}
\end{equation}
with
$$\chi=-3\frac{\dot a}{a}=-\frac{2t}{1+t^2}.$$
acquiring the maximal values of $\pm 1$ at $t=\mp 1$ when $\dot\chi=2\frac{t^2-1}{(t^2+1)^2}=0$. A rather curious feature of notations is that  $\chi$ (proportional to the Hubble parameter) reaches a maximum value of $\sqrt{\frac{3}{2}}\chi_m$, and not $\chi_m$ as the name would suggest.

One can check that (\ref{bounce}) is indeed a solution by direct substitution into (\ref{rRay}). There is, however, a very peculiar point. Namely, we have
$$1-\chi^2=\left(\frac{t^2-1}{t^2+1}\right)^2,$$
and the solution comes only by assuming
$$\sqrt{1-\chi^2}=\frac{t^2-1}{t^2+1}$$
which features two changes of the branch for the square root function, though required by smoothness with respect to time. 

At $t=0$ the branch with $f(0)=f^{\prime\prime}(0)=-\frac43$ is used, and the bounce does happen. In the Ref. \cite{CMcosmo} this change of branch was automatically accomodated because the modified Friedman equation (\ref{modFried}) is obtained by squaring the equation (\ref{Fried1}). Another direct way to observe this would be look at the modified Friedamnn equation \eqref{modFried}. As noted in \cite{CMcosmo}, the expression for the constant $C$ (corresponding to a contribution of the mimetic matter) can be given (by substituting the solution (\ref{lambdasol}) into the Friedmann equation (\ref{Fried1})) as
\begin{eqnarray}\label{Csoln}
   \frac{C}{a^3}= \frac{1}{3}\chi^2 +f -\chi f'\,,
\end{eqnarray}
and therefore the expression for the energy density (assuming only mimetic dark matter) can be written as
\begin{eqnarray}\label{epsilon}
    \frac{3}{2}\, \epsilon = 1- \sqrt{1- \chi^2}\,.
\end{eqnarray}
It is easy to see from \eqref{modFried} that, at the bounce, $\epsilon = \epsilon_m = 4/3$ (the last equality comes from our convention of choosing $\chi_m^2 = 2/3$). Then, at this point, we need to choose the negative square root on the r.h.s. of \eqref{epsilon} in order to satisfy the equation. 

This is the crucial point which we wish to convey in this article: $\chi=0$ gives rise to two different regimes, one corresponding to the bounce while the other to Minkowski spacetime. As evidenced from \eqref{Csoln} above, we need the function $f$ to correspond to two different values for these two limits, thereby choosing different branches of this multi-valued function.

The change of branch is controlled by a subtle mechanism. The second derivative diverges $|f^{\prime\prime}|\to\infty$ when $\chi\to\pm 1$. It makes $\dot\chi\to 0$ keeping all terms in Friedmann equations finite. Therefore, $1-\chi^2\propto (\delta t)^2+{\mathcal O}((\delta t)^3)$ around $t=\pm 1$ enabling us to consistently choose $\sqrt{1-\chi^2}\propto \delta t+{\mathcal O}((\delta t)^2)$. It is not easy to imagine this singularity of $f^{\prime\prime}$ being resolved in some more fundamental theory since it is absolutely essential for changing the branch without digressing to complex numbers. It remains to be seen whether such mechanism can work stably for generic solutions, with the argument of $f$ reaching the branching hypersurface with precisely one transversal derivative vanishing at each point.

Note also that the function $f$ is infinitely branched\footnote{It would be interesting to investigate how much does the mechanism rely on this feature. If not to insist on the Born-Infeld analogy, a term $(1-\chi^2)^{3/2}$ would give the same singular behaviour of $f^{\prime\prime}$ at $\chi^2\to 1$ with bounded $f^{\prime}$. Then additional ${\mathcal O}(1)$ and ${\mathcal O}(\chi^2)$ terms would easily allow to restore GR in the principal branch, while ${\mathcal O}(\chi^4)$ and higher corrections can be used for adjusting big $\chi$ behaviour, if needed.}, due to $\arcsin$. However, the branches of the $\arcsin$ function come in pairs of the form "$\arcsin \to \pi - \arcsin$" which correspond to the binary branching of the square root (which is also of importance for keeping $f^{\prime}$ finite at $\chi\to 1$ on all branches), and the pairs are repeated with periodicity of $2\pi$. The latter makes only shifts to $f(\chi)$ by $-\frac{4\pi}{3} \chi$ which are not important (unless we need to keep track of boundary terms) since $\chi\equiv\square\phi$.

We conclude this section by remarking on the inevitable nature of working with such multi-valued functions to achieve bounce in such models. It was noted in passing in \cite{CMcosmo} that perhaps a change of variables such as $\psi=\sin(\chi)$ would end our need to work with a multi-valued $f$. Although it is true that one can then think of a single-valued function as follows
\begin{eqnarray}
	f(\psi) = 1 + \frac{1}{2}\sin(\psi)^2 - \psi \sin(\psi) - \cos(\psi)\,,
\end{eqnarray}
it is still the variable $\chi$ which is related to the basic fields of the model. Determining the new variable $\psi$ in terms of the physical field $\chi$ would of course require to use a multi-valued function (inverse sine) again\footnote{The crucial point here is that $f$ turns out to be a function of not just $\sin(\psi)$, but also $\cos(\psi)$. Therefore, $\sin(\psi) \rightarrow 0$ as $\chi \rightarrow 0$ is not sufficient to fully describe the function and indeed, $\cos(\psi)$ goes to $\pm 1$ in the two regimes of interest.}.

\section{\label{sec:bey}Beyond the simplest function}
One of the most alluring features of the modified Friedmann equation (\ref{modFried}) and the bouncing solution (\ref{bounce}) lies in their remarkable simplicity. The analytical complexity increases a lot, even when the mere coefficient in front of $f$ is changed. Let us briefly consider what happens if
$$f(\chi)=-\left(\frac23 +2\beta\right)\cdot \left(1+\frac12 \chi^2-\chi \arcsin\chi- \sqrt{1-\chi^2}\right)$$
with $\beta\neq 0$.
Substituting this function, together with solution (\ref{lambdasol}), into eq. (\ref{Fried1}) we get
\begin{equation}
\label{newrel}
\epsilon+\beta\chi^2+\left(\frac23 +2\beta\right)\sqrt{1-\chi^2}=\frac23 +2\beta,
\end{equation}
and in right hand side of the modified Friedmann equation, instead of $\epsilon\left(1-\frac34 \epsilon\right)$ , we find a solution of the quadratic equation for $3H^2=\frac13 \chi^2$:
$$3H^2=-\frac{\frac13 \left(\frac23 +2\beta\right)+\epsilon\beta}{3\beta^2}\left(1\pm\sqrt{1+\frac{\beta^2\left(2\left(\frac23 +2\beta\right)\epsilon-\epsilon^2\right)}{\left(\frac13 \left(\frac23 +2\beta\right)+\epsilon\beta\right)^2}}\right)$$
where $\epsilon=\frac{C}{a^3}$.

At small $\beta$ one of the roots (if positive) corresponds to a very fast (super-Planckian) universe, while another one tends to the previous $\epsilon\left(1-\frac34 \epsilon\right)$ solution as $\beta\to 0$. In the latter case, the gross cosmological features apparently remain the same. The bounce occurs at the critical density of $\epsilon_{cr}=2\left(\frac23 +2\beta\right)$ where $H=0$ featuring a parabolic ($\delta a\propto t^2 +{\mathcal O}(t^3)$) behaviour in time. Somewhere between $\epsilon=0$ and $\epsilon=\epsilon_{cr}$ the function $\chi(\epsilon)$ reaches its maximal value at which $\dot\chi=\frac{d\chi}{d\epsilon}\dot\epsilon=0$. Of course, by differentiating the relation (\ref{newrel}) with respect to $\epsilon$, we see that it happens ($\frac{d\chi}{d\epsilon}=0$) at $\chi=\pm1$ which justifies changing the branch precisely as in the previous Section.

The analytic structure of eq. (\ref{newrel}) does not lend much support to an idea of deep relations between the Chamseddine-Mukhanov bounce and LQC. The chosen function (\ref{rf}) is admittedly a very natural choice. The square root structure is dictated by the very essence of the bouncing mechanism, while the $\arcsin$ is the simplest solution for making $f^{\prime}$ finite, and other terms ensure avoiding low-energy corrections to GR. However, the overall coefficient cannot be fixed by any reasonable motivations. Yet, as we have seen, changing it drives the model away from the simple LQC equation.

\section{\label{sec:ons}On similarity with loop quantum cosmology}
Recently, this model has also gained quite a bit of attention due to its similarities with loop quantum cosmology \cite{LQC1, LQC2}. The modified Friedmann equation, as derived in \eqref{modFried}, is the same as the `effective' Friedmann equation coming from LQC. This correspondence is, however, limited to the minisuperspace case as has already been noted in \cite{LQC1,LQC3}. Let us first point out a possible pitfall in such a correspondence if extended to the full theory. As mentioned towards the end of Section \ref{BranchChange}, there are additional boundary terms which show up due to the function changing branches which are  not important for the purpose of deriving the modified Einstein equations. However, the starting point for full loop quantum gravity is the Holst action, where one adds a topological term to the usual Einstein-Hilbert action, with the coefficient identified as the Immirzi parameter (a fundamental parameter within the theory). Although this term does not affect the classical behaviour, it results in an inequivalent quantum theory (as compared to say, usual Wheeler-deWitt quantizations). Thus, the effect of additional terms can be crucial for the full quantum theory in LQG and must, therefore, be handled with suitable care.

Another interesting point is that one does not have to take care of such branch cuts while dealing with LQC and this can seem to be at loggerheads to our construction above. The reason is that, in going from the CM model to LQC, one has to define one of the new variables through a canonical transformation involving the inverse sine function. As is shown in both \cite{LQC1, LQC2}, one needs to define variables such as
\begin{eqnarray}\label{LQCtrans}
	b \sim \arcsin(\chi)\,,
\end{eqnarray}
and we have to be precise about the multi-valued function. It is important that one chooses the branch where, roughly speaking, $b$ corresponds to $\pi$ and not $0$ at the point of bounce when $\chi$ goes to zero. This is why the final form of the LQC Hamiltonian, as also derived from the CM model, is thus free from such ambiguities. 

This is an important conceptual point which has not been mentioned in previous works so far \cite{LQC1,LQC2}. For instance, in \cite{LQC2}, the authors fix some of the co-efficients in the equation corresponding to \eqref{LQCtrans} (Eqn. (3.28) in \cite{LQC2}) by using the weak-gravity limit and taking the principal branch of the arcsin function. This is evidently correct. However, as already mentioned several times over, while approaching the bounce one needs to focus on a different branch of the arcsin function to get the correct values there, which needs to be emphasized.

Since one can derive the same effective Hamiltonian (constraint) as in LQC, we can carry over some of the conclusion of the latter theory to the CM model. For instance, near the bounce, $\chi$ starts increasing from zero, and $\dot\chi$ is positive till $\chi$ reaches its maximum value before starting to decrease again. This phase, increase of $\chi$, or equivalently $H$, from zero to a maximum symbolizes an era of superinflation which is, of course, nothing specific to LQC but rather a tautological consequence of having a bounce. In LQC, it has been shown that this superinflation era is extremely short-lived (time duration equals $1$ in units of our eq. (\ref{bounce}) i.e. Planck scale for LQC). In the CM model, the limiting curvature is set at a scale below Planck scales which gives longer absolute time $\sim\frac{1}{\sqrt{\epsilon_m}}$, however it still corresponds to less than one e-fold of expansion. Nevertheless, as was discussed for LQC in Ref. \cite{AS}, a superinflationary period might make initial conditions for subsequent inflation more favourable, and this conclusion can seemingly be carried over to the CM model.

\section{Discussion}

As we have seen, the modifed mimetic model of Refs. \cite{CMcosmo, CMBH} requires a subtle use of multi-valued functions in order to avoid the initial singularity of the Friedmann universe. The obstacle for working with a single-valued function is precisely the condition (\ref{condition}) for reproducing GR. The better we screen weak gravity from unwanted modifications, the less probable is a bounce if we measure depatures from GR solely by the current value of $\square\phi$. This conclusion is very generic. In anisotropic cosmologies, there still remains open a possibility of a separated bounce when some dimensions bounce while others rapidly change. Another option is to make the argument of the function $f$ better correspond to the curvature at the bouncing transition. One could, for example, use $f(\square^2\phi)$ but it would be hard to keep such models stable.

Conceptually, the CM mechanism is very peculiar. For using such a model, it is not enough to know the values of all dynamical fields at a given Cauchy hypersurface; rather, one must also know at which branch the model currently is. It has to be taken with great care when considering relations to LQC. As we have mentioned before, the change of branch within real numbers essentially employs the singularity in $f^{\prime\prime}$ making it hard to envisage a more fundamental and regular theory behind the model. On the other hand, LQC might be seen as such a resolution for one particular function $f$, modulo the tricky issue of the boundary terms. 

We end this paper by questioning whether using such a mechanism can be viewed as opening a Pandora's box in a rather interesting way? Suppose, one's favourite $f(R)$ model for, say, quantum Universe creation,  contradicts experimental tests of GR. Can one go look for a function with two branches, for the newborn and the adult Universe separately, so as to satisfy observational constraints? We leave these interesting questions for future investigations.

\section*{Acknowledgments}
The authors are grateful to V. Mukhanov for illuminating discussions. This work was supported by the Korean Ministry of Education, Science and Technology, Gyeongsangbuk-Do and Pohang City for Independent Junior Research Groups at the Asia Pacific Center for Theoretical Physics.

\end{document}